\documentclass[pre,twocolumn,letterpaper,showpacs,superscriptaddress,floatfix]{revtex4}
\usepackage{graphicx,latexsym,amsfonts,amsmath,amssymb,bm}

\begin{document}
\title{Anisotropic Diffusion Limited Aggregation}
\date{\today}
\author{M. N. Popescu}
\email{popescu@mf.mpg.de}
\affiliation{Max-Planck-Institut f\"ur Metallforschung,
Heisenbergstr. 3,D-70569 Stuttgart, Germany}
\affiliation{Institut f\"ur Theoretische und Angewandte Physik,
Universit\"at Stuttgart, Pfaffenwaldring 57, D-70569 Stuttgart, Germany}
\author{H.G.E Hentschel}
\email{phshgeh@physics.emory.edu}
\affiliation{Department of Physics, Emory University, Atlanta, GA, 30322, USA}
\author{F. Family}
\email{phyff@emory.edu}
\affiliation{Department of Physics, Emory University, Atlanta, GA, 30322, USA}

\begin{abstract}
Using stochastic conformal mappings  we study the effects of anisotropic perturbations
on diffusion limited aggregation (DLA) in two dimensions. The harmonic measure of the
growth probability for DLA can be conformally mapped onto a constant measure on a
unit circle. Here we map $m$ preferred directions for growth of angular width $\sigma$
to a distribution on the unit circle which is a periodic function with $m$ peaks
in $[-\pi, \pi)$ such that the width $\sigma$ of each peak scales as
$\sigma \sim 1/\sqrt{k}$, where $k$ defines the ``strength'' of anisotropy along
any of the $m$ chosen directions. The two parameters $(m,k)$ map out a parameter
space of perturbations that  allows a continuous transition from DLA
(for $m=0$ or $k=0$) to  $m$ needle-like fingers as $k \to \infty$. We show that
at fixed $m$ the effective fractal dimension of the clusters $D(m,k)$  obtained from
mass-radius scaling decreases with increasing $k$ from $D_{DLA} \simeq 1.71$ to a
value bounded from below by $D_{min} = 3/2$. Scaling arguments suggest a specific
form for the dependence of the fractal dimension $D(m,k)$ on $k$ for large $k$, form
which compares favorably with numerical results.
\end{abstract}

\pacs{05.45.Df, 61.43.Hv}

\maketitle

\section{Introduction}

Nonequilibrium growth models leading naturally to self-organized fractal structures,
such as diffusion limited aggregation (DLA)~\cite{WS81}, have received great interest
in the recent years not only due to their relevance for various physical processes,
for example dielectric breakdown~\cite{NPW84}, electrochemical
deposition~\cite{BB84,MSHHS84}, and two-fluid Laplacian flow~\cite{P84}, but
also because such harmonic growth leads naturally to one of the most interesting
multifractal distributions found in nature~\cite{HP83,HMP86}.

A powerful method for studying such two dimensional growth processes is the iterated
stochastic conformal mapping~\cite{HL98,H97,DHOSS99}, which has already been
successfully applied to generate and analyze DLA~\cite{DHOSS99,DP00} and
Laplacian~\cite{BDLP00} growth patterns in two dimensions. This has opened the road
to address many important questions related to pattern formation in DLA, such as the
structure of the multifractal spectrum of DLA~\cite{JLMP02}, and provided the
first definite answers for how the hottest tips and the coldest fjords grow. Other
topics that can be investigated using iterated conformal maps include the pinning
transition in Laplacian growth~\cite{HPF02}, the difference between Hele-Shaw flows
and DLA~\cite{HLP02}, as well as new topics such as the scaling of fracture surfaces
formed during quasistatic cracking~\cite{BHLP02}.

One of the important questions addressed soon after the original discovery of DLA by
Witten and Sander~\cite{WS81} was that of the effect of the intrinsic anisotropy in
lattice models on the shape and fractal dimension of the asymptotic
aggregates~\cite{BBRT85_86,M88,EMPZ89_90,AACR91,JS95}. For two dimensional growth, it
was shown that the result of such anisotropy in the microscopic attachment probability
leads to clusters which asymptotically have the symmetry of the underlying lattice
(following the argument in Ref~\cite{BBRT85_86}, this actually holds for $m \leq 6$,
where $m$ represents the coordination number of the lattice), and the fractal dimension
of the resulting aggregate asymptotically approaches $3/2$. These results have also
been confirmed in recent work which used iterated stochastic conformal mapping
techniques to grow the clusters~\cite{SL01}.

In the present work we use iterated stochastic conformal mapping techniques
to study DLA with $m$ preferred directions for growth. Although this naturally leads to
anisotropic clusters, the present model is fundamentally different from the
previous studies on lattice anisotropy. Our model is rather related to the
existence of a large scale imposed $m$-fold symmetry whose strength can be tuned.
Specifically let us consider the case when the harmonic measure for DLA is
weighted at angle $\psi$ between the seed and the location for growth by a
term $W(\psi;m,\sigma )$, where $\sigma$ specifies the angular width of the
preferred direction. Such a weighting
$W(\psi ; m,\sigma )\sim \exp[-\beta {\cal H}(\psi ; m,\sigma )]$
could be due to an imposed external field or to growth on a surface which has an
$m$ fold symmetry. An example would be dendritic growth in a strip~\cite{AACR91}
which can be argued to lie in the $m=1$ or $m=2$ universality class, the anisotropy
increasing as the strip is narrowed.

The organization of the paper is as follows. In Section II we describe how we use
conformal mapping methods together with an angle dependent probability for
growth $P(\theta;m,k)$ to study a model corresponding to a real space weighting
$W(\psi;m,\sigma )$. Here $\theta$ is the angle parametrising the unit circle
to which the boundary of the growing cluster is conformally mapped, $m$ is the number
of the privileged directions, and $k$ is an appropriate measure for the ``strength''
of the anisotropy. In Section III we present results for the morphology
of the resulting patterns as a function of $m$ and $k$, and we derive using
scaling arguments the effective fractal dimension $D(m,k)$ of the emerging
clusters. We conclude with a discussion of the results in Section IV.

\section{Model and Theoretical Background}

In the DLA model proposed by Witten and Sander~\cite{WS81} the growth of a
cluster from a seed placed at the origin proceeds by irreversible attachment
of random walkers released from infinity (in practice, from far away from the
cluster's boundary). Thus the probability $P(s)$ for growth at any point $s$
along the cluster boundary of total length $L$ is a harmonic measure and can be
written $P(s) = |(\nabla V)(s)|/\int_0^L |(\nabla V)(s')| ds'$, where $V({\bf r})$
is a potential which outside the cluster obeys Laplace's equation $\nabla^2 V = 0$
subject to the boundary conditions $V = 0$ on the (evolving) boundary of the
cluster and $V \sim \ln r$ as $r \to \infty$ (corresponding to a uniform source
of particles far away from the cluster). In two dimensions, this formulation
as a potential problem has been recently exploited for studying the time development
of DLA based on conformal mapping techniques ~\cite{HL98,DHOSS99}.

\begin{figure}[!htb]

\begin{minipage}[c]{.95\columnwidth}
\hspace*{.3in}\includegraphics[width=.62\textwidth]{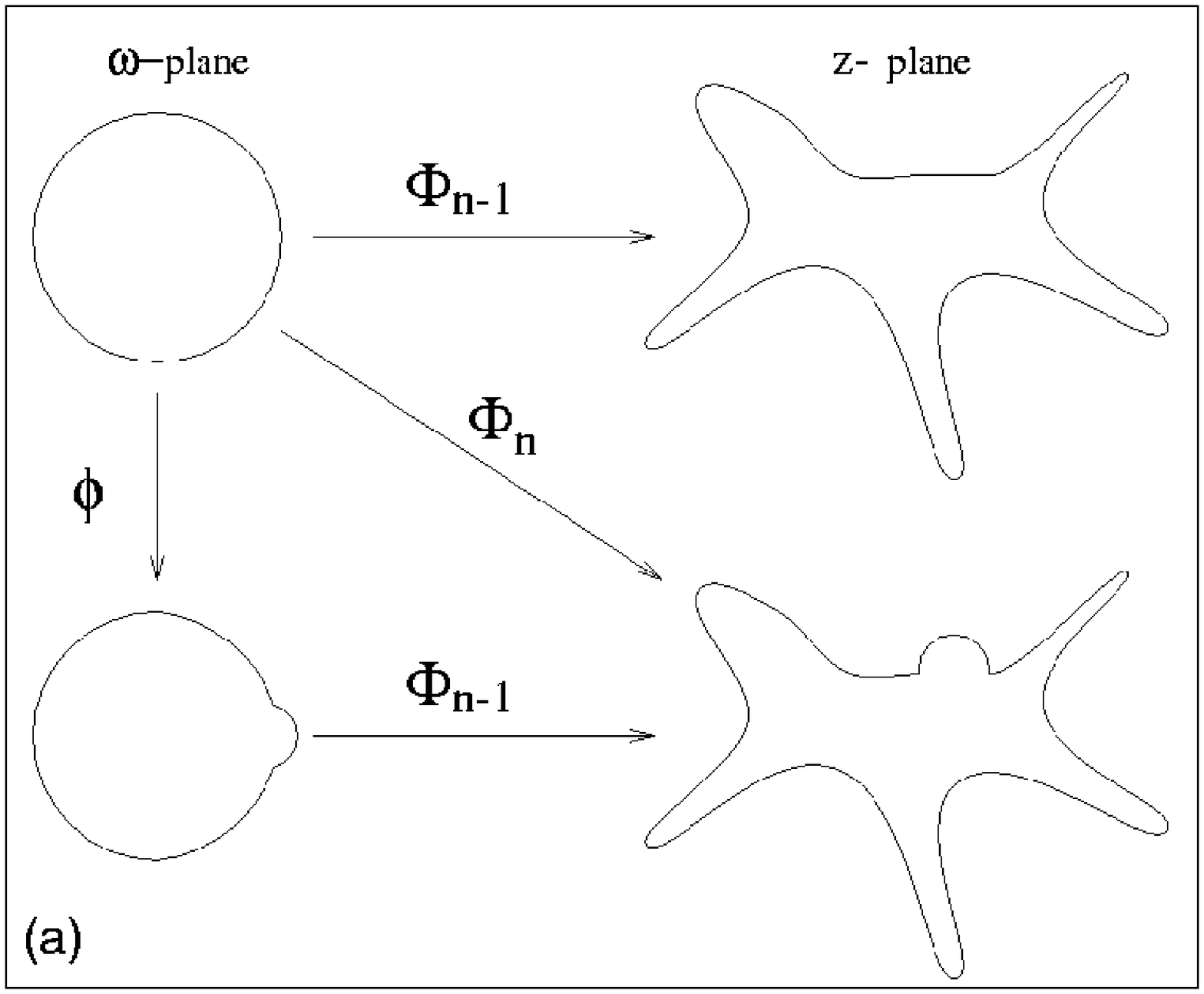}
\end{minipage}%

\begin{minipage}[c]{.95\columnwidth}
\vspace*{.2in}
\includegraphics[width=.7\textwidth]{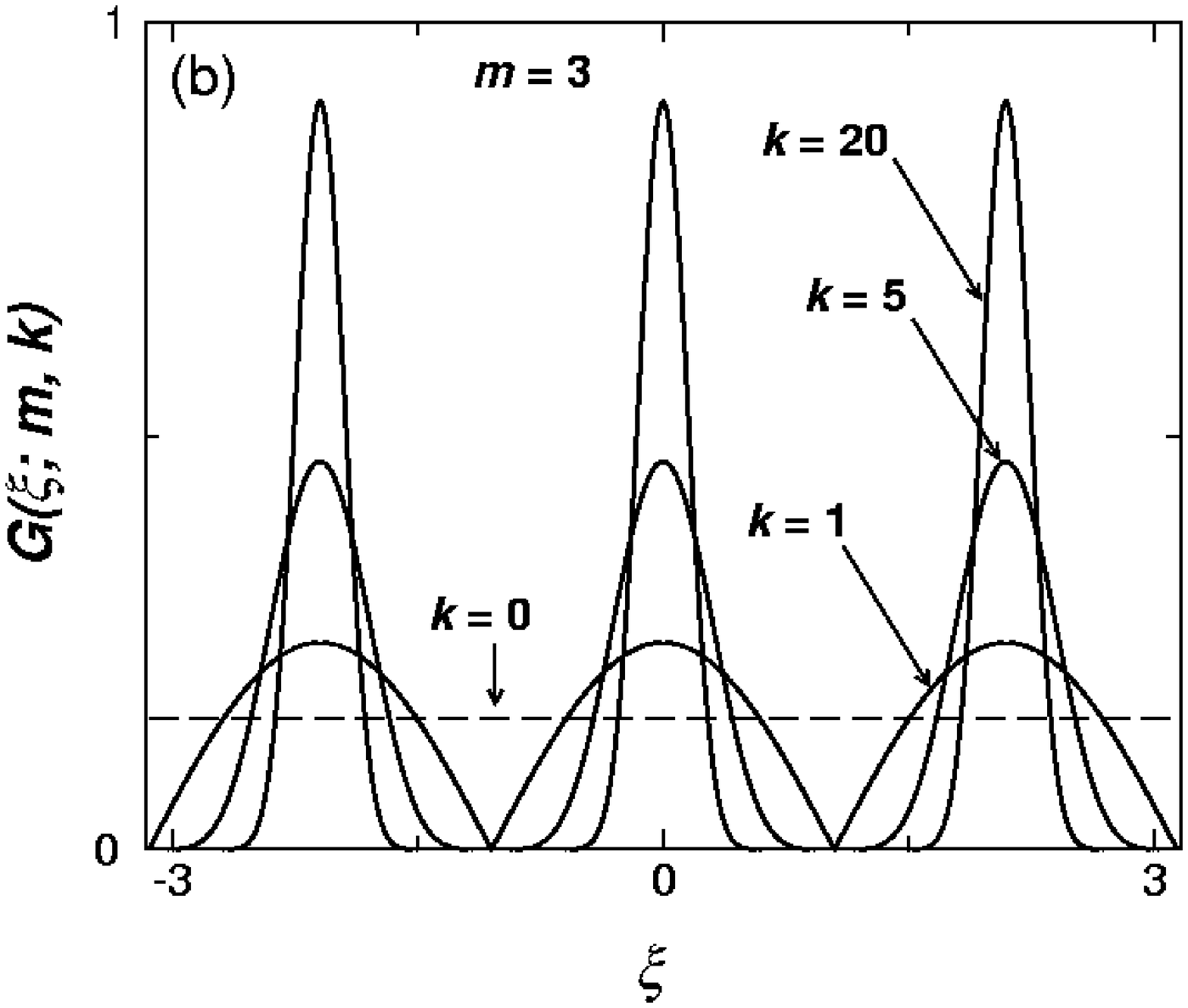}
\end{minipage}%

\caption{\label{fig1}
(a) Diagrammatic representation of the mappings $\Phi$ and $\phi$.
(b) Change in shape of the probability distribution $G(\xi;m,k)$ with
increasing $k$  for $m = 3$.}
\end{figure}

As discussed in detail in ~\cite{HL98,DHOSS99}, the basic idea is to follow the
evolution of the conformal mapping $\Phi^{(n)}(\omega )$ of the exterior of the unit
circle in a mathematical $\omega$--plane onto the complement of the cluster of $n$
particles in the physical $z$--plane rather than directly the evolution of the
cluster's boundary. The equation of motion for $\Phi^{(n)}(\omega)$ is determined
recursively (see  Fig.~\ref{fig1}(a)). With an initial condition corresponding to
the unit circle in the physical plane $\Phi^{(0)}(\omega) = \omega$, the process of
adding a new ``particle'' of constant shape and linear scale $\sqrt{\lambda_0}$ to
the cluster of $(n-1)$ ``particles'' at a position $s$ chosen according to the
harmonic measure is performed using an elementary mapping
$\phi_{\lambda, \theta}(\omega )$
\begin{eqnarray}
&&\phi_{\lambda,0}(\omega) =
\omega^{1-a} \left\{ \frac{(1+ \lambda)}{2\omega}(1+\omega)\right.
\nonumber\\
&& \left.\times \left [ 1+\omega+\omega \left( 1+\frac{1}{\omega^2} -
\frac{2}{\omega} \frac{1-\lambda}
{1+ \lambda} \right) ^{1/2} \right] -1 \right \} ^a \nonumber\\
&&\phi_{\lambda,\theta} (\omega)
= e^{i \theta} \phi_{\lambda,0}(e^{-i \theta} \omega) \,,
\label{eq-f}
\end{eqnarray}
which conformally maps the unit circle to the unit circle with a bump of size
$\sqrt{\lambda}$ localized at the angular position $\theta$ ~\cite{HL98}.
The parameter $a$ describes the shape of the elementary mapping; following
the analysis in ~\cite{DHOSS99}, we have used $a=0.66$ throughout this paper
as we believe the large scale asymptotic properties will not be affected by
the microscopic shape of the added bump.
As shown diagrammatically in Fig.~\ref{fig1}(a), the recursive dynamics can
than be represented as iterations of the elementary bump map
$\phi_{\lambda_{n},\theta_{n}}(\omega)$, resulting in the convolution representation
of the conformal map $z = \Phi^{(n)}(\omega)$ at the $n$th stage of growth as
\begin{equation}
\Phi^{(n)}(\omega) = \phi_{\lambda_1,\theta_{1}}
\circ\phi_{\lambda_2,\theta_{2}}\circ\dots\circ
\phi_{\lambda_n,\theta_{n}}(\omega)\, ,
\label{comp}
\end{equation}
where the angle $\theta_n \in [-\pi, \pi)$ at step $n$ is randomly chosen
because the harmonic measure on the real cluster translates to a uniform
measure on the unit circle in the mathematical plane, i.e,
\begin{equation}
P(s) ds = \frac{d\theta}{2 \pi}.
\label{prob}
\end{equation}
Eq.~\ref{prob} is crucial to the successful implementation of the iterated
conformal method as the highly nontrivial harmonic measure in the physical
plane becomes uniform in the mathematical plane. Finally,
\begin{equation}
\lambda_{n} = \frac{\lambda_0}{|{\Phi^{(n-1)}}' (e^{i \theta_n})|^2}
\label{lambda}
\end{equation}
is required in order to ensure that the size of the bump in the physical
$z$ plane is $\sqrt{\lambda_0}$. We note that in the composition Eq.~\ref{comp} the
order of iterations is inverted -- the last point of the trajectory is the
inner argument, therefore the transition from $\Phi^{(n)}(\omega )$ to
$\Phi^{(n+1)}(\omega )$ is achieved by composing the $n$ former maps
Eq.~\ref{comp} starting from a different point.

Consider the case where the existence of $m$ preferred directions in
physical space modulates the harmonic measure at any point $s$ on the boundary
by a probability
$P(\psi(s); m,\sigma ) = W(\psi(s);m,\sigma )/\int_0^L W(\psi(s');m,\sigma)ds'$.
Here, $\psi(s)$ is the angle parameterisation of the cluster boundary in the
physical space, $W(\psi;m,\sigma )$ is the modulating weight, and the $m$-fold
periodicity implies~$P(\psi~+~2\pi/m;m,\sigma) = P(\psi;m,\sigma)$.
The important question is \textit{if} the weighting $W(\psi;m,\sigma )$
in the real space may be represented in the form of a modulation of the constant
measure in the mathematical plane $P_{math}(\theta) = d\theta/2 \pi$.
Because the angle $\psi$ is not invariant under the conformal map
$z = \Phi^{(n)}(\omega)$, an answer to the question above is not straightforward.
Considering an ensemble of clusters generated under the influence of the same
modulation $P(\psi; m,\sigma )$, for each cluster of $n$ particles $\psi$
maps onto a different $\theta_n(\psi)$, where
$\exp{i\psi}=\Phi^{(n)}(\exp{i\theta_n})/|\Phi^{(n)}(\exp{i\theta_n})|$.
It is reasonable to assume that averaging over the many patterns above
an asymptotically ($n \to \infty$) average scale invariant pattern will appear.
For DLA, i.e., in the absence of modulation, this pattern is a circle; in the
general case, an $m$ fold periodic pattern with the same symmetry as the
modulation is expected to appear. Therefore, we expect
$\langle \theta_n (\psi) \rangle = f_m(\psi )$, where $\langle \dots \rangle$
denotes an average over clusters, with $f_m(\psi )$ independent of $n$ and
satisfying $f_m(\psi + 2\pi/m) = f_m(\psi )+2\pi/m$ due to the symmetry of the
resulting averaged pattern and to the fact that $f_m$ is an angle. In principle,
$f_m$ is defined up to an additive constant; this can be fixed by specifically
requiring that at the pattern fingertips $j = 0,1,2,...m-1$ one has
$f_m(2\pi j/m) = 2\pi j/m$. It then follows that the modulated probability
distribution on the unit circle leading to such $m-$fold symmetry patterns obeys
\begin{equation}
P_{math}(\theta ) = P(f_m^{-1}(\theta ); m,\sigma ) df_m^{-1}(\theta ) /d\theta
\label{relation}
\end{equation}
Thus, we can see that for this case $P_{math}(\theta )$ is itself an $m$-fold
periodic function on the unit circle with peaks at the preferred directions
$\theta_k = 2\pi k/m$.

In this paper, rather than attempting to derive a specific $P_{math}(\theta )$
using Eq.~\ref{relation},  we shall directly assume such an $m$-fold periodic
measure on the unit circle (which based on the arguments above is expected to
lead to an $m-$fold symmetry weighting function $W$ in the physical space)
and study the clusters created using the choice
$P_{math}(\theta) = G(\theta;m,k) d\theta$, where the parameter
$k \sim 1/\sigma$ is an appropriate measure of the angular width of the
preferred direction in the physical plane. The angle-dependent probability
distribution on the unit
circle $G(\theta ;m,k)$ will be normalized such that
$\int_{-\pi}^{\pi} d\theta \, G(\theta;m,k) = 1$. Such a distribution biases
the choice of the location $\theta$, and thus $s$, where growth occurs as
follows. At step $n$, the point $s$ for the attempt of growth is chosen, as before,
based on the harmonic measure, i.e., one chooses points $\theta_n \in [-\pi,\pi)$
on the unit circle with uniform distribution. But growth at $s$ is only allowed
with a probability $G(\theta_n;m,k)$. If the attempt is rejected, then the
previous sequence is repeated until a successful trial occurs. We note
that obviously $G(\theta;m,k) = const$ corresponds to usual DLA, while an
explicit dependence on $\theta$ models the existence of privileged directions.

Because at this stage we are interested in the general features of such
a model of anisotropic growth, and not in trying to model a specific
physical system, we will make the simple choice
\begin{eqnarray}
&&G(\theta;m,k) = \frac{1}{C(k)}
\left|\cos\left(\frac{m}{2} \theta\right) \right|^k,
\nonumber\\
&&\theta \in {\left[ \right.} -\pi, \pi \left. \right),
~m \in \mathbb{N}~, ~k \in \mathbb{R}_+ ~,
\label{prob_G}
\end{eqnarray}
where
\begin{equation}
C(k) = \frac{4\,{\sqrt{\pi }}\,\Gamma(\frac{3}{2} + \frac{k}{2})}
{\left( 1 + k \right) \,\Gamma(1 + \frac{k}{2})} \nonumber
\label{constant}
\end{equation}
is the normalization constant (note that it does not depend on $m$).

It is easy to see that $G(\theta;m,k)$ defined above has all the key properties
required: for $m > 0$ it is a periodic function of $\theta$ of principal period
$2 \pi/m$, and thus the number $m$ of peaks of $G(\theta;m,k)$ in
$\left[-\pi, \pi\right)$ corresponds to the number of privileged directions;
obviously, both $k=0$ independent of $m$ and $m = 0$ independent of $k$
correspond to isotropic DLA growth. The exponent $k > 0$ is a measure for the
``strength" of selectivity, i.e., the larger $k$, the narrower and
higher peaks of $G(\theta;m,k)$. Therefore, the pair $(m,k)$ defines a two
dimensional parameter space for the analytic study of applied anisotropy to DLA.
Though the choice given by Eq.~\ref{prob_G} is arbitrary, we believe that
because of universality the key features will be independent of the specific
form of the function $G(\theta;m,k)$.

\section{Results and Discussion}

\begin{figure}[!htb]

\begin{minipage}[c]{.95\columnwidth}
\includegraphics[width=.9\textwidth]{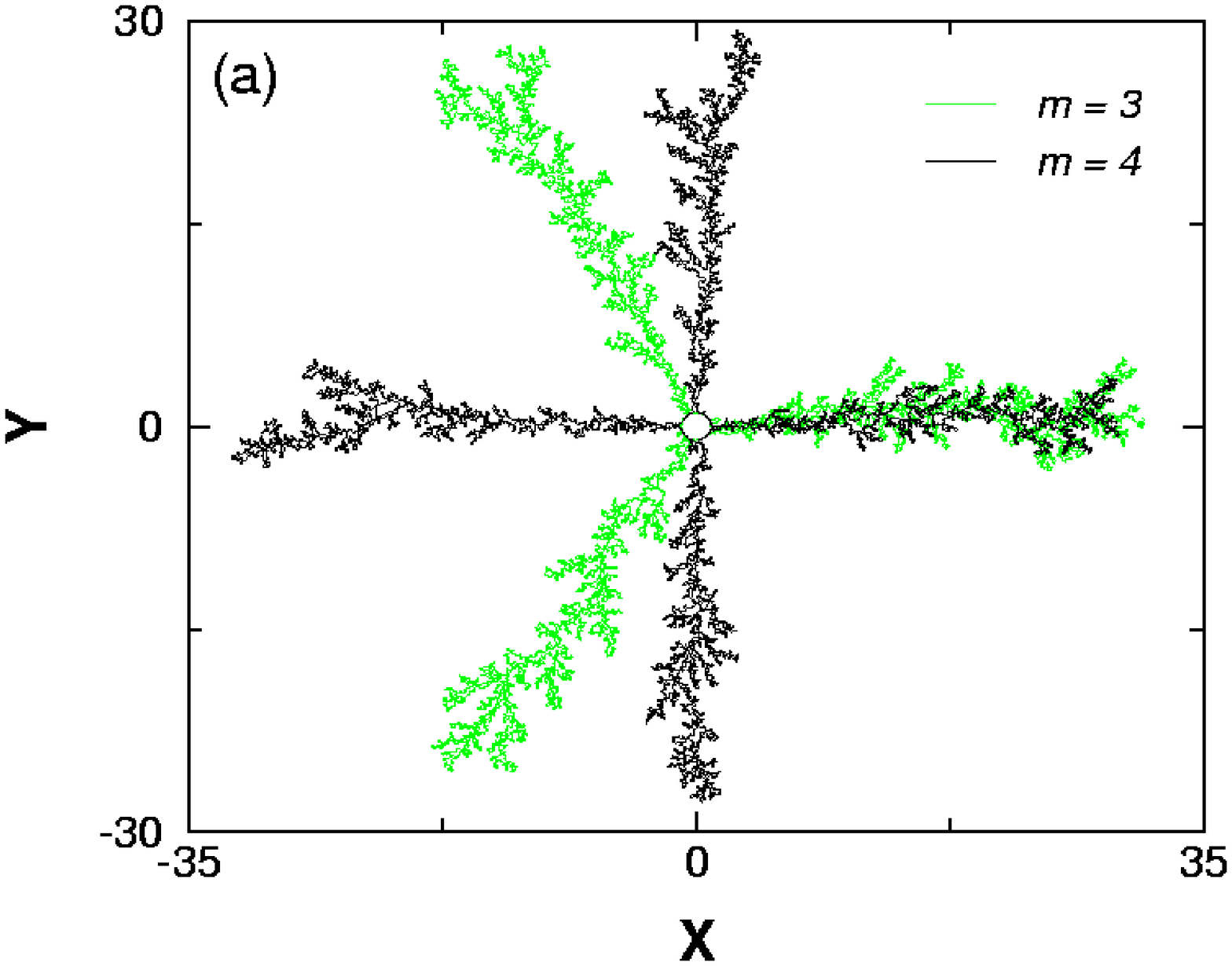}
\end{minipage}%

\begin{minipage}[c]{.95\columnwidth}
\includegraphics[width=.9\textwidth]{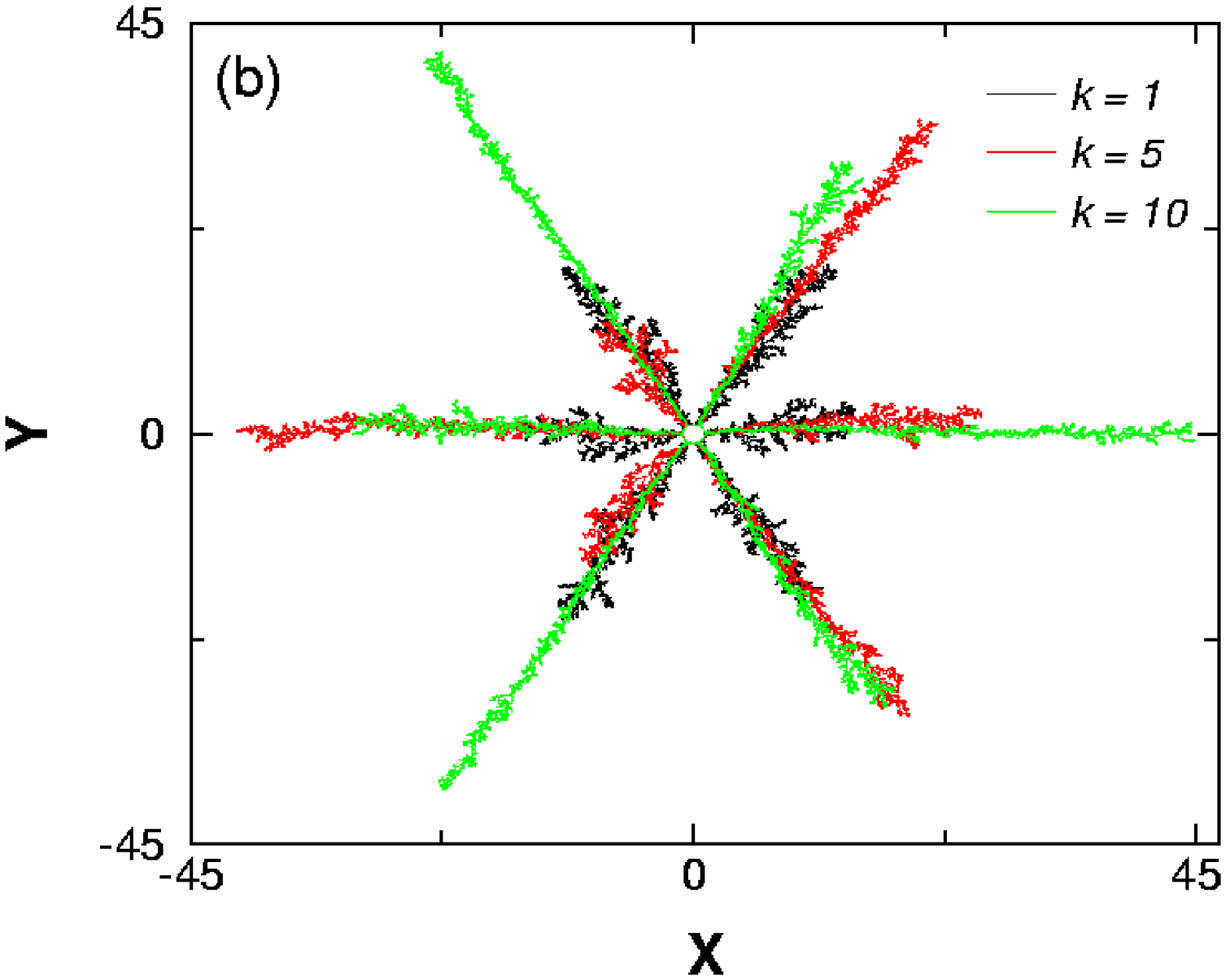}
\end{minipage}%

\caption{\label{fig2}
Typical clusters (size $N=20000$) grown with (a) $m = 3, 4$, fixed $k = 3$,
and (b) fixed $m = 6$, but different values for $k$, $k = 1, 5, 10$, respectively.}
\end{figure}

The model described in Sec.~II was simulated as follows. The parameter
$\lambda_0 = 10^{-3}$ was fixed because it is just setting the microscopic
area of an added ``particle''. For fixed $m$ and $k$ the growth step $n$
proceeds by selecting at random (uniform probability) an angle
$\theta \in [-\pi,\pi)$, then comparing $G(\theta;m,k)$ with a random number $r$
uniformly distributed in $[0, 1/C(k))$; if $r < G(\theta;m,k)$, then
$\theta_n = \theta$, $\lambda_n$ follows from Eq.~\ref{lambda}, and the new
map $\Phi_n$ follows from Eq.~\ref{comp}; if not, the previous sequence is
repeated until a successful trial occurs. All the averages mentioned below
were done over $100$ clusters grown up to size $N = 20000$. An example of
clusters with different symmetries (i.e., different values $m$)
grown in this manner is shown in Fig.~\ref{fig2}(a), while Fig.~\ref{fig2}(b)
depicts the morphology with increasing anisotropy ``strength" (i.e.,
different values $k$ at fixed $m$).

It can be seen that the bias introduced by the distribution $G(\theta;m,k)$ is
indeed producing clusters with the corresponding $m-$ fold symmetry and that it
is very effective: even for small values $k$, the clusters in
Fig.~\ref{fig2}(a) show a clear 3-fold, respectively 4-fold, symmetry.
Increasing $k$ (the strength of the anisotropy) leads to a significant
reduction in the branched structure of the cluster, thus in the thickness
of the surviving branches, as shown in Fig.~\ref{fig2}(b). Similar results have
been obtained for all the values $2 \leq m \leq 7$ and $1 \leq k \leq 80$
that have been tested.

\begin{figure}[!htb]

\begin{minipage}[c]{.95\columnwidth}
\includegraphics[width=.9\textwidth]{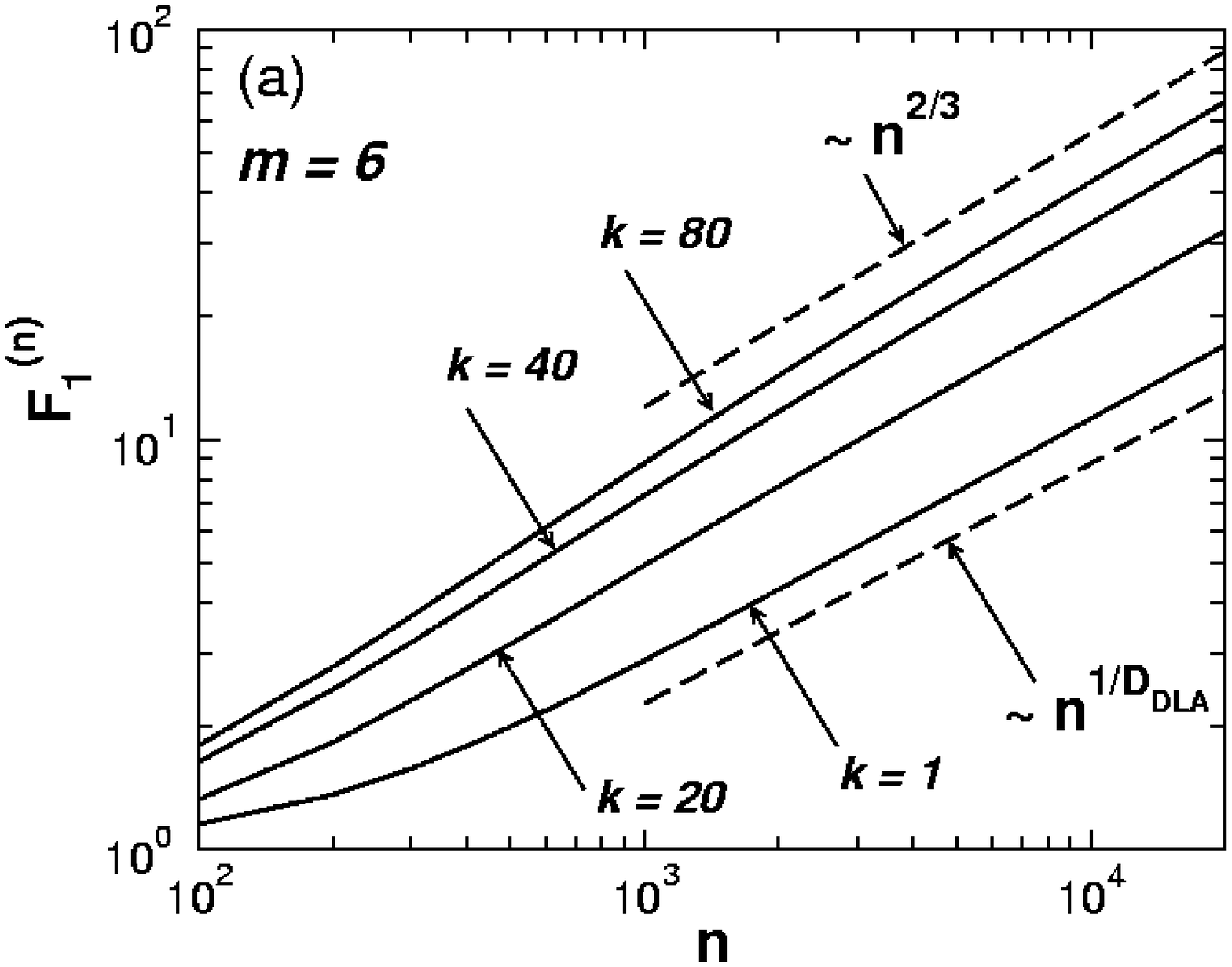}
\end{minipage}%

\begin{minipage}[c]{.95\columnwidth}
\hspace{.1in}\includegraphics[width=.9\textwidth]{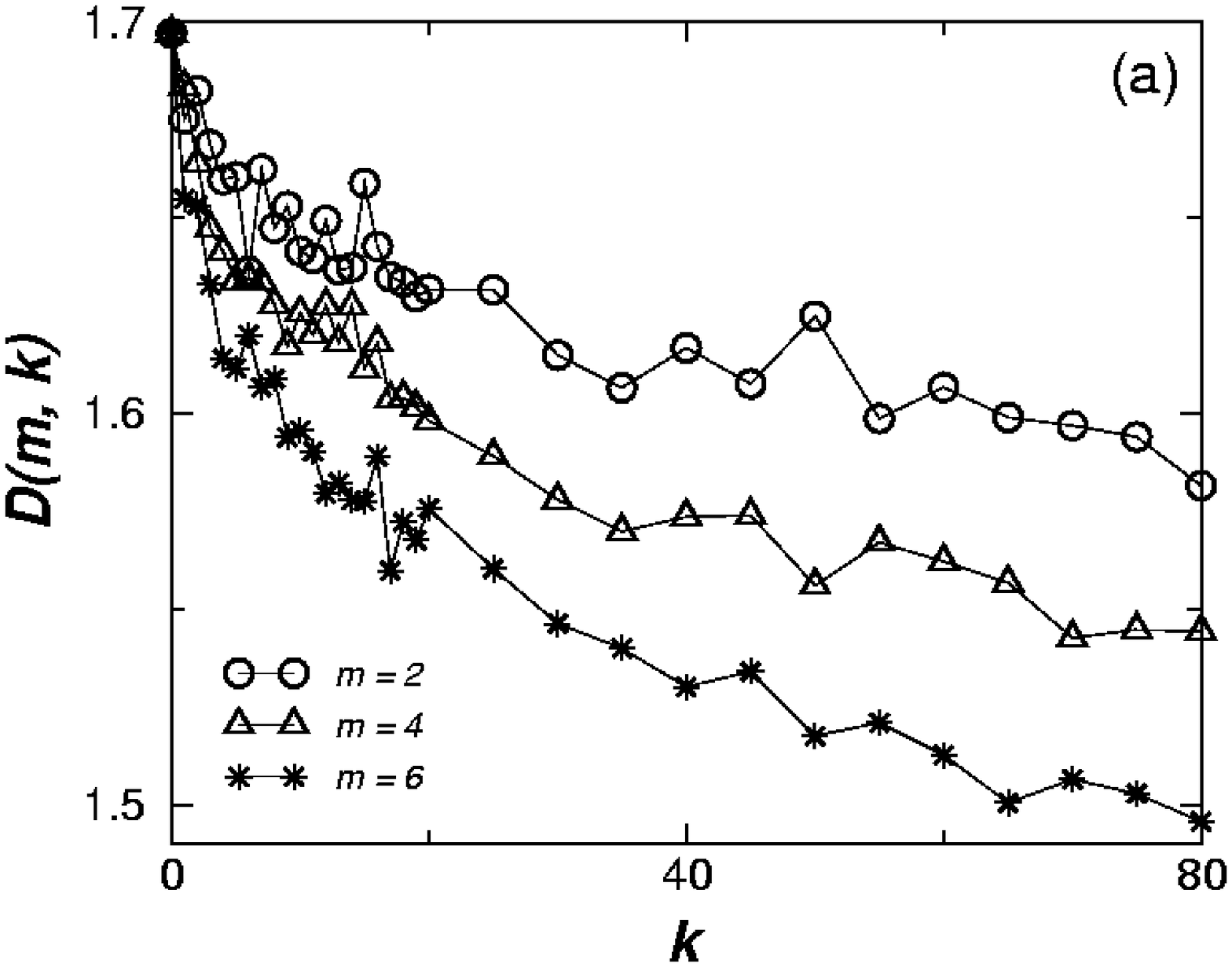}
\end{minipage}%

\caption{\label{fig3}
(a) Average $F_1^{(n)}$ as a function of $n$ for clusters grown with
$k = 1, 10, 40, \textrm{and } 80$, respectively, and fixed $m = 6$
(log-log plot). Also shown (dashed lines) are both the limit case
$F_1^{(n)} \sim n^{1/D_{DLA}}$ (DLA cluster), where
$D_{DLA} = 1.71$, and the proposed lower bound for anisotropic
DLA growth ~\cite{BBRT85_86} $F_1^{(n)} \sim~ n^{1/D_{min}}$,
where $D_{min} = 3/2$.
(b) The effective fractal dimension $D(m,k)$, obtained from
$F_1^{(n)} \sim n^{1/D(m,k)}$, as a function of $k$ at fixed $m$.
The points represent the measured values, the lines are just
a guide to the eye.}
\end{figure}

The results shown in Fig.~\ref{fig2} suggest that the resultant patterns have a
fractal morphology that depends on $m$ and $k$, and in order to
characterize these shapes we will focus on the effective fractal dimension
$D(m,k)$ obtained from the mass-radius scaling. Following the arguments
in Ref.~~\cite{DHOSS99}, the coefficient
$F^{(n)}_1 = \Pi_{i=1}^n (1+ \lambda_i)^a$ in the Laurent
expansion of $\Phi^{(n)}$,
\begin{equation}
\Phi^{(n)}(\omega) = F^{(n)}_{1} \omega + F^{(n)}_{0} + F^{(n)}_{-1}\omega^{-1} +
F^{(n)}_{-2}\omega^{-2} + \dots ~,
\label{eq-laurent-F}
\end{equation}
is a typical length-scale of the cluster; thus, a natural choice for the
radius of the $n$-particle cluster is $R \sim F_1^{(n)}$. Assuming that for
$n >> 1$ a scaling law of the form
\begin{equation}
\label{F1_scaling}
F_1^{(n)} \sim n^{1/D(m,k)}
\end{equation}
is found, the effective fractal dimension of the cluster can be extracted
from a power law fit to the numerical data. We note in passing that
this scaling law was used in Ref.~~\cite{DHOSS99} as a very convenient
way to measure the fractal dimension of the growing DLA cluster.
As we have anticipated, for all the values $m$ and $k$ the numerical
results for the average coefficient $F_1^{(n)}$ show clear power-law
dependence on the size $n$, an example being shown in Fig.~\ref{fig3}(a),
and the results $D(m,k)$ obtained from the power-law fit to the
data in the range $n \geq 10^3$ are shown in Fig.~\ref{fig3}(b).

It can be seen that $D(m,k)$ decreases with increasing $k$ at fixed $m$
(and with increasing $m$ at fixed $k$), and there is a certain tendency
for saturation at large $k$. We note that, as expected,
$D(m,0) \simeq D_{DLA}$ and that the curves $D(m,k)$ are all above the
expected lower limit $D_{min} = 3/2$ ~\cite{BBRT85_86}. An exception is
the case $m=7$ (results not shown), where for large $k$ the values
$D(7,k >> 1) \simeq 1.45$ are somewhat below $D_{min}$, but this
is most probably due to either insufficient statistics (too few clusters),
as suggested also by the noisiness of the $D(m,k)$ curves,
or to the fact that in this particular case the size $N = 20000$ is not
sufficient to obtain an asymptotic cluster.

\begin{figure}[!htb]

\begin{minipage}[c]{.95\columnwidth}
\includegraphics[width=.9\textwidth]{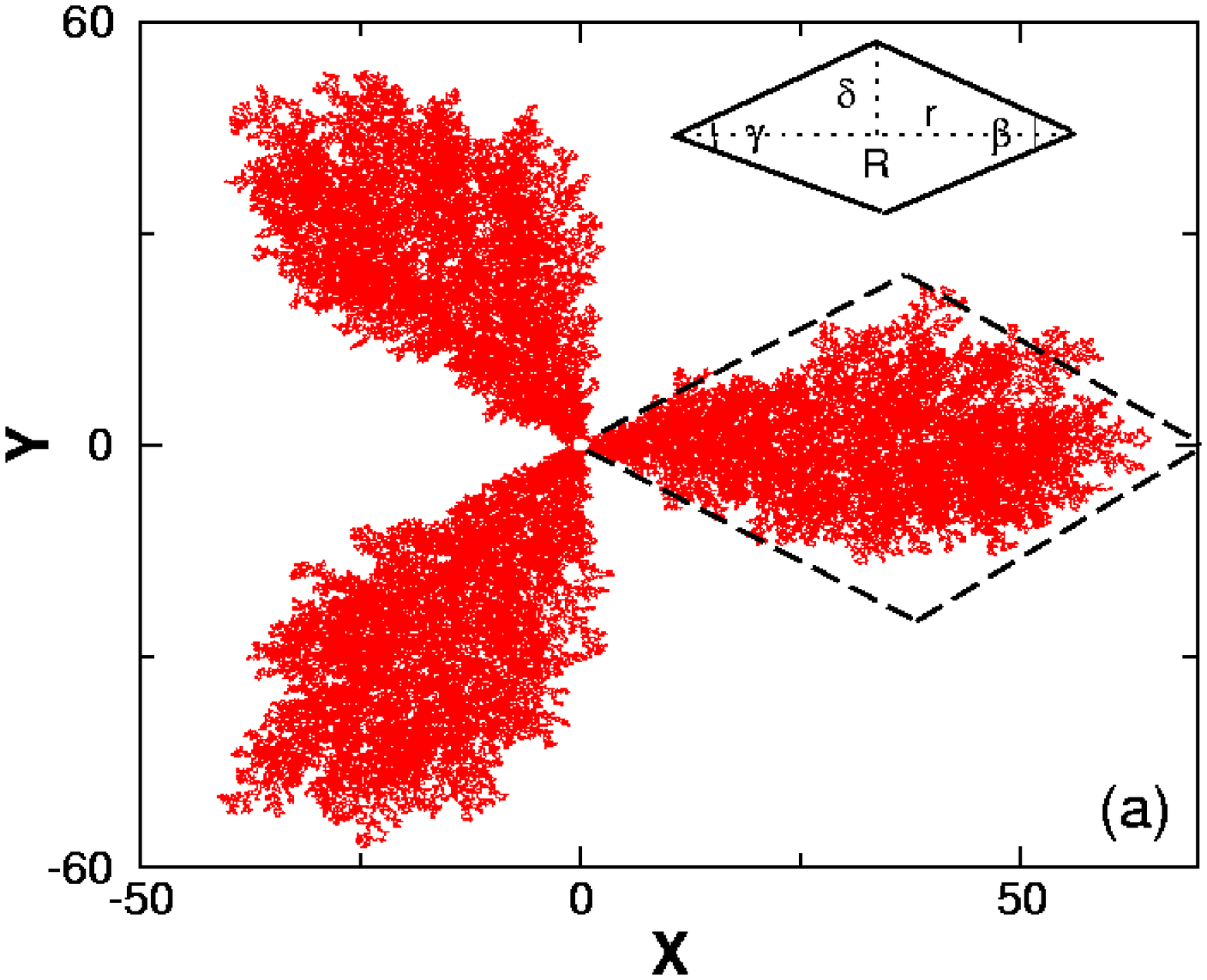}
\end{minipage}%

\begin{minipage}[c]{.95\columnwidth}
\includegraphics[width=.92\textwidth]{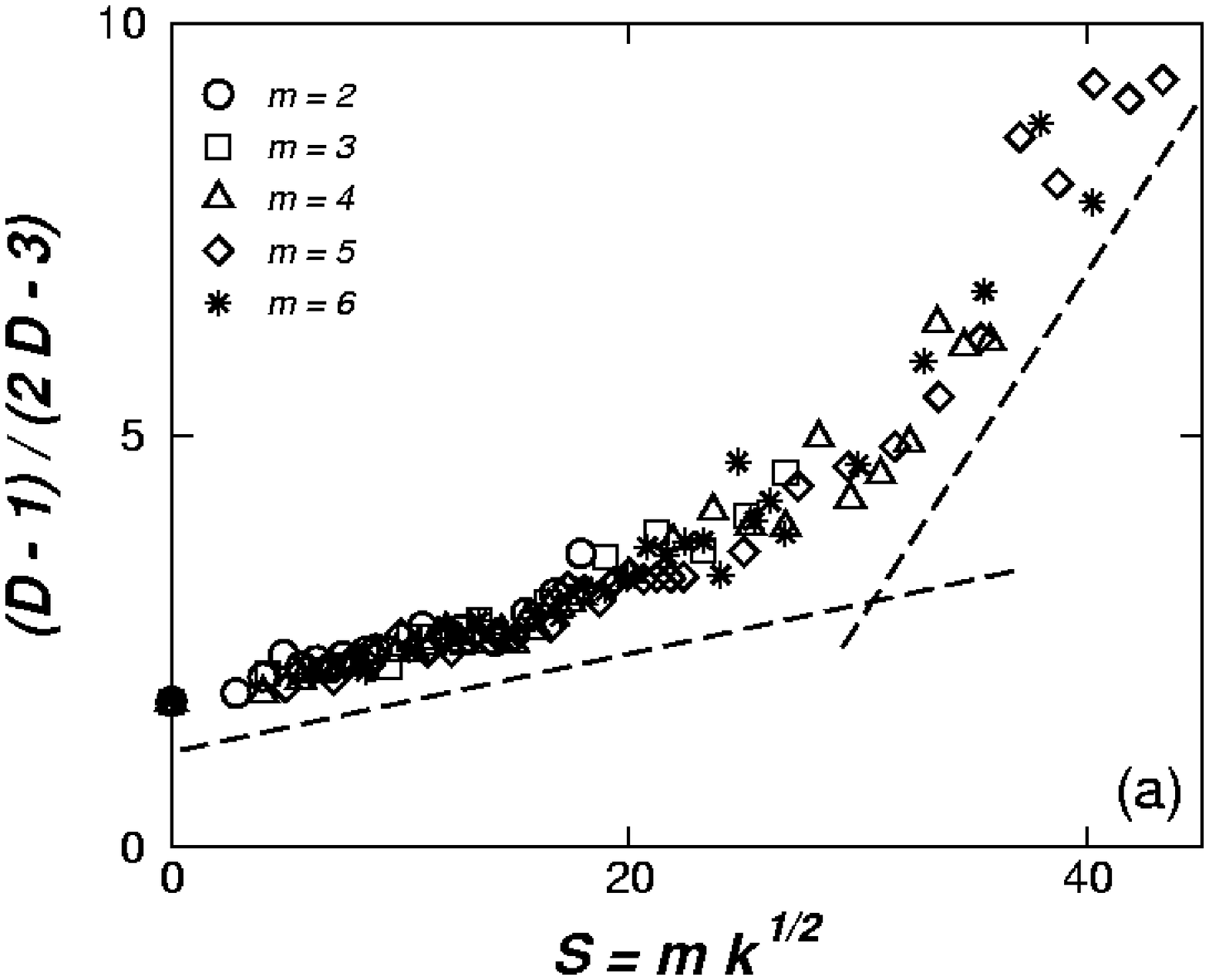}
\end{minipage}%

\caption{\label{fig4}
(a). Superposition of $10$ different clusters of size $N = 10^5$ grown with
the same $m = 3$ and $k = 1$ but different sequences of random numbers. The
dotted diamond around the arm centered at $\psi = 0$ shows the approximation
for the envelope of one arm of the cluster, and the drawing in the upper right
corner shows schematically the geometry of the diamond.
(b) Numerical results for $\frac{D-1}{2 D - 3}$ as a function of the scaling
variable $S = m \sqrt{k}$. The dashed lines are just a guide to the eye
for the linear behavior in the range $S >> 1$ and, respectively, $S \gtrsim 1$.}
\end{figure}

In order to understand these results theoretically we will use a simple
argument, following Ref.~~\cite{BBRT85_86}, based on the assumptions that
(a) for large $k$ the growth of the cluster occurs mainly at the tips of
the $m$ principal branches, and (b) the envelope of the \textit{average}
cluster can be approximated by $m$ diamond shaped polygons, like the
one shown in Fig.~\ref{fig4}(a), of opening angles $\gamma$ and $\beta$
(in general, these angles depend on both $m$ and $k$) and with edges of
lengths in the order of $R$.

Under these assumptions, the rate of growth can be written as ~\cite{BBRT85_86}
\begin{equation}
\label{a_rate}
dN/dR \sim R^{\pi/(2 \pi - \beta)}.
\end{equation}
Because the LHS of Eq.~\ref{a_rate} is $LHS \sim R^{D(m,k)-1}$, once the angle
$\beta (m,k)$ is known $D(m,k)$ can be determined from
\begin{equation}
\label{D_beta}
D(m,k) = 1 + \frac{\pi}{2 \pi - \beta (m,k)}
\end{equation}
Simple geometry (see the schematic drawing in the top right corner of
Fig.~\ref{fig4}(a)) allows one to write (under the assumption that the angles
$\gamma (m,k)$ and $\beta (m,k)$ are small -- which is certainly true for large
$k$ and $m$),
\begin{equation}
\label{beta_gama}
\gamma = \frac{\delta}{R-r},~\beta = \frac{\delta}{r}
~\Rightarrow ~\beta = \frac{R-r}{r}\,\gamma
\end{equation}
On the other hand, the opening angle $\gamma (m,k)$ is obviously fixed by the
decay of the probability for growth $G(\theta;m,k)$, and therefore can be
estimated as the width of the peak of the distribution. Working with
the peak centered at $\theta = 0$, assuming large $k$ and small $\gamma$,
the width at half peak  probability
is given by $1/2 \approx \left[1-\left(m \gamma/4\right)^2\right]^k
\approx 1 - k \left(m\gamma/4\right)^2$. Thus, from Eq.~\ref{beta_gama},
\begin{equation}
\label{beta}
\gamma (m,k) \approx \frac{2}{m \sqrt{k}}
~\Rightarrow \beta (m,k) \approx \frac{2(R-r)}{r \, m \sqrt{k}} =
\frac{C_1}{m \sqrt{k}}
\end{equation}
where $C_1$ is  a constant of ${\cal O}(1)$, independent of $m$, $k$, or the size
$N$ of the cluster. Combining Eqs.~\ref{D_beta} and \ref{beta}, one thus obtains
the following scaling relation for the fractal dimension
\begin{equation}
\label{scaling}
\frac{D(m,k)-1}{2 D(m,k) - 3} = \frac{\pi}{C_1} m \sqrt{k}.
\end{equation}
Note that it is immediately apparent from Eq.~\ref{scaling} that $D_{min}= 3/2$,
while for large $k$ the ratio $(D-1)/(2D-3)$ should be a linear function of
the product $S = m \sqrt{k}$  \textit{only}, provided $S >> 1$. This prediction
can be tested against the numerical results. As shown in Fig.~\ref{fig4}(b), the
data collapse is excellent and the function is indeed linear when $S>>1$,
confirming our theory. Surprisingly, the scaling predicted by Eq.~\ref{scaling}
seems to hold  down to quite small values of $k$, although these results are
beyond the scope of our scaling arguments.

\section{Conclusions}

Using iterated stochastic conformal maps, we have studied the patterns
emerging from a model of anisotropic (in the sense of the existence of
privileged radial directions for growth) diffusion limited aggregation
in two dimensions. In our model, the anisotropy was introduced via a
probability distribution for growth with a number $m$ of peaks in
$[-\pi, \pi)$, the width of a peak (giving the ``strength'' of anisotropy)
being a tunable parameter that allows a continuous transition from isotropic
DLA growth to anisotropic clusters. We have shown numerical evidence that
at fixed $m$ the effective fractal dimension of the clusters $D(m,k)$ obtained
from the mass-radius scaling decreases with $k$ from $D_{DLA}$ to values
bounded from below by $D_{min} = 3/2$. Using simple approximations (supported by
numerical results) for the envelope of the cluster and general scaling
arguments, we have derived a scaling law involving $D(m,k)$ and successfully
tested it against numerical results.

Although the model we have proposed is somewhat artificial, it has the
advantage that it seems to capture most of the general features of an
anisotropic growth process while it is still simple enough to allow an
analytical treatment (to a certain degree). Finally, we note here that a
system for which the proposed geometry may be easily experimentally achieved
is the growth of bacterial colonies. For such a case, the radial anisotropy
can be experimentally obtained through the addition of nutrients along the
privileged directions, and controlled through the excess concentration of
nutrients along these directions in respect to the rest of the substrate. This
would allow a direct testing of all our numerical and analytical predictions.

\begin{acknowledgments}

This work has been supported by the Petroleum Research Fund. One of us (MNP)
would like to thank the Physics Department at Emory University for the very
warm hospitality during the visit when part of this work has been done.

\end{acknowledgments}


\begin{thebibliography}{}

\bibitem{WS81}
T.A. Witten Jr. and L.M. Sander, Phys. Rev. Lett. {\bf 47},1400 (1981).

\bibitem{NPW84}
L. Niemeyer, L. Pietronero, and H.J. Wiesmann, Phys. Rev. Lett. {\bf 52},1033 (1984).

\bibitem{BB84}
R.M. Brady and R.C. Ball, Nature {\bf 309}, 225 (1984).

\bibitem{MSHHS84}
M. Matshushita, M. Sano, Y. Hayakawa, H. Honjo, and Y. Sawada,
Phys. Rev. Lett. {\bf 53},286 (1984).

\bibitem{P84}
L. Paterson, Phys. Rev. Lett. {\bf 52},1621 (1984).

\bibitem{HP83}
H.G.E. Hentschel and I. Procaccia, Physica  D {\bf 8}, 435 (1983).

\bibitem{HMP86}
T.C. Halsey, P. Meakin and I. Procaccia, Phys. Rev. Lett. {\bf 56}, 854 (1986).

\bibitem{HL98}
M.B. Hastings and L.S. Levitov, Physica D {\bf 116}, 244 (1998).

\bibitem{H97}
M.B. Hastings, Phys.Rev.E {\bf 55}, 135 (1997).

\bibitem{DHOSS99}
B. Davidovich, H.G.E. Hentschel, Z. Olami, I.Procaccia, L.M. Sander, and E.
Somfai, Phys. Rev. E {\bf 59}, 1368 (1999).

\bibitem{DP00}
B. Davidovich and I. Procaccia, Phys. Rev. Lett. {\bf 85}, 3608 (2000).

\bibitem{BDLP00}
F. Barra, B. Davidovitch, A. Levermann, and I. Procaccia,
Phys. Rev. Lett. {\bf 87}, 134501 (2001).

\bibitem{JLMP02}
M.H. Jensen, A. Levermann, J. Mathiesen, and I. Procaccia,
Phys. Rev. E {\bf 65}, 046109 (2002).

\bibitem{HPF02}
H.G.E. Hentschel, M.N. Popescu, and F. Family,
Phys. Rev. E {\bf 65}, 036141 (2002).

\bibitem{HLP02}
H.G.E. Hentschel, A. Levermann, and I. Procaccia,
Phys. Rev. E {\bf 66}, 016308 (2002).

\bibitem{BHLP02}
F. Barra, H.G.E. Hentschel, A. Levermann, and I. Procaccia,
Phys. Rev. E65, 045101 (2002).

\bibitem{BBRT85_86}
R.C. Ball, R.M. Brady, G. Rossi, and B.R. Thompson
Phys. Rev. Lett. {\bf 55}, 1406 (1985); R.C. Ball,
Physica {\bf 140A}, 62 (1986).

\bibitem{M88}
P. Meakin, in {\it Phase Transitions and Critical Phenomena},
edited by C. Domb and J. Lebowitz (Academic, New York, 1988), Vol. 12.

\bibitem{EMPZ89_90}
J.P. Eckmann, P. Meakin, I. Procaccia, and R. Zeitak,
Phys. Rev. A {\bf 39}, 3185 (1989); {\it ibid},
Phys. Rev. Lett. {\bf 65}, 52 (1990).

\bibitem{AACR91}
A. Arnedo, F. Argoul, Y. Couder, and M. Rabaud,
Phys. Rev. Lett. {\bf 66}, 2332 (1991).

\bibitem{JS95}
B.K. Johnson and R.F. Sekerka, Phys. Rev. E {\bf 52}, 6404 (1995).

\bibitem{SL01}
M.G. Stepanov and L.S. Levitov, Phys. Rev. E {\bf 63}, 061102 (2001).

\bibitem{TS85} L. Turkevich and H. Sher, Phys. Rev. Lett. {\bf 55}, 1026 (1985).

\end{thebibliography}
\end{document}